\documentclass{WileyMSP-template}

\usepackage{amsmath}
\usepackage{amssymb}
\usepackage{lineno}
\usepackage{tabularx}
\usepackage{makecell}
\usepackage{graphicx}
\usepackage{booktabs}
\usepackage{subcaption}
\usepackage{bm}
\usepackage{hyperref}
\usepackage{siunitx}
\usepackage{url}
\usepackage{graphicx}
\usepackage{float}
\newfloat{scheme}{htbp}{los}
\floatname{scheme}{Scheme}
\floatname{chart}{Chart}
\newfloat{graph}{htbp}{loh}
\usepackage{chemformula} 
\usepackage[version = 4]{mhchem} 

\begin{document}

\pagestyle{fancy}
\rhead{\includegraphics[width=2.5cm]{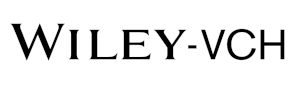}}

\title{End-to-End Inverse Designed Single-Layered Metasurface for High-Compression Snapshot RGB-Achromatic Full-Stokes Polarization Imaging}

\maketitle


\author{Xingyu Chai}
\author{Jirong Bao}
\author{Haining Yang}
\author{Mengdi Sun*}


\dedication{}

\begin{affiliations}
Xingyu Chai$^1$, Jirong Bao$^1$, Prof. Haining Yang$^1$, Prof. Mengdi Sun$^{1,2}$*\\
$^1$ School of Electronic Science and Engineering, Southeast University, Nanjing 211189, China\\
$^2$ Jiangsu Key Laboratory for Undersea Communications and Sensing, China\\
Email Address: smd@seu.edu.cn
\end{affiliations}



\begin{abstract}

Snapshot full-Stokes polarimetry across multiple wavelengths remains challenging because conventional architectures rely on multiplexed measurements and bulky optics. We present an end-to-end inverse designed single-layered metasurface that reconstructs RGB full-Stokes images from a snapshot sensor measurement. A metasurface modeled by a multilayer perceptron (MLP) is employed to encode the full-Stokes polarization information. The system jointly optimizes a differentiable metasurface frontend with a U-Net backend. On a real-world dataset, our design achieves a high compression ratio of 12 for snapshot RGB-achromatic polarization imaging using a single-layered metasurface. These results show that end-to-end optical-digital co-design enables high-compression snapshot full-Stokes polarization imaging with a compact footprint.

\end{abstract}


\section{Introduction}

Polarization imaging provides information beyond conventional intensity imaging by probing the vector nature of light. In particular, full-Stokes measurements detect both linear and circular polarization components and therefore capture richer information on surface roughness\cite{walter1998polarization}, depolarization\cite{Lu96}, multiple scattering\cite{david2000polarization, ren2025single, xie2025reciprocal}, and chiral or anisotropic responses\cite{wang2022anisotropic, song2023polarization, zhou2022polarization, neubrech2020reconfigurable}, with applications ranging from scene understanding and remote sensing to imaging in challenging environments\cite{chen2024polarization, wolff1997polarization, lei2021polarimetric}. However, most full-Stokes imaging systems still rely on spatially or temporally multiplexed systems with polarizers\cite{liu2025}, wave plates\cite{Peinado:13}, or division-of-focal-plane architectures\cite{Gruev:10}, which limits their use in compact platforms\cite{ma2025integrated, pierangeli2023single}. Furthermore, temporally multiplexed systems require multiple exposures and thus cannot capture fast-moving objects or dynamic scenes\cite{9363502, Wagadarikar}, while the spatially multiplexed systems can hardly capture the full information of the objects in a snapshot measurement under limited detector-pixel resources, especially at high compression ratios. 

Metasurfaces are planar subwavelength nanopatterns and enable complementary metal-oxide-semiconductor (CMOS) compatibility, system miniaturization and massive production. They can tailor polarization response within an ultrathin form factor\cite{khorasaninejad2016metalenses, chen2024broadband}. This capability has motivated a broad range of meta-optic polarization and spectro-polarimetric devices\cite{junzhuo2023recent, balthasarmueller2017metasurface, Thrane2026}, but its application in highly compressed full-Stokes imaging is still limited due to the lack of the ability of snapshot imaging. Computational imaging provides an alternative route to snapshot reconstruction from compressed measurements, but its performance is inevitably constrained by the complex and bulky frontend optical elements\cite{pierangeli2024deep}.

To address these limitations, computational imaging has increasingly shifted from post hoc reconstruction to optical-digital co-design, in which the optical frontend and the reconstruction backend are optimized jointly for a target inference task\cite{baek2021single, roquescarmes2025metaoptic, frch2025computational}. Recent work has shown that meta-optics can serve not merely as miniature replacements for conventional optics, but as task-specific physical encoders trained together with the reconstruction backend\cite{xia2024joint, arya2024end, lin2022end, lin2021end, fisher2025end}. This perspective is particularly compelling for full-Stokes imaging, where the polarization-multiplexed snapshot measurement is intrinsically compressive and the reconstruction quality depends as much on the optical encoding as on the digital decoding.

Here we present an end-to-end meta-optic framework for snapshot RGB-achromatic full-Stokes polarization imaging. Our system combines a differentiable meta-optics frontend and a U-Net reconstruction backend. To minimize the computational cost and make the pipeline compatible with gradient-based optimization, we use a differentiable multilayer perceptron (MLP)-based surrogate model to predict the polarization response of the meta-atoms instead of running expensive numerical electromagnetic solvers. The metasurface geometry, the object/image distances, and the neural-network weights are simultaneously adjusted during optimization. On a real-world spectro-polarimetric dataset\cite{jeon2024spectral}, our design achieves a high compression ratio of 4 for monochromatic imaging with 27.06 dB peak signal-to-noise ratio (PSNR) and 0.7172 structural similarity index measure (SSIM), and an even high higher compression ratio of 12 for RGB-achromatic imaging with 23.35 dB PSNR and 0.5643 SSIM. These results show that end-to-end co-design opens a new avenue for ultra-compact, snapshot, full-Stokes polarization imaging systems. 

\section{Method}

Our framework combines a differentiable encoder with a neural decoder for snapshot full-Stokes polarization imaging. The optical frontend is based on a single-layered polarization-sensitive metasurface, and the computational backend is a U-Net that reconstructs the full-Stokes image cube from a snapshot sensor measurement. The meta-optic layer is only $\approx 500$ $\mathrm{\mu}$m thick, far thinner than conventional optical systems with millimeter or centimeter scale sizes. To enable efficient end-to-end optimization, the metasurface is modeled through an MLP-based surrogate model trained on full-wave electromagnetic simulation results. The system is optimized by jointly updating the metasurface geometry, the object/image distances, and the neural-network weights. In Section \ref{sec2.1}, we detail the design process of the MLP-based surrogate model and the optical frontend within the proposed framework, as shown in \textbf{Figure \ref{fig1}} and \textbf{Figure \ref{fig2}}. In Sections \ref{sec2.2} and \ref{sec2.3}, we further detail the computational backend and the end-to-end optimization pipeline, as shown in Figure \ref{fig2}. 

\subsection{Differentiable optical frontend}

\label{sec2.1}
\begin{figure}[t] 
    \centering
    \includegraphics[width=0.8\textwidth]{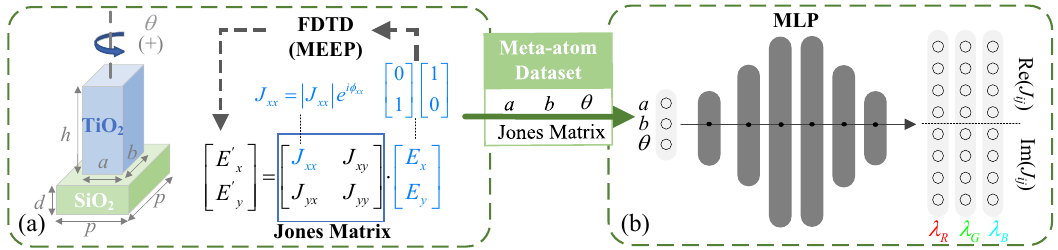}
    \caption{The design process of the MLP-based surrogate model for the metasurface. (a) The schematic of the unit cell geometry and the corresponding optical responses. (b) The training process of the MLP-based surrogate model.}
    \label{fig1}
\end{figure}

\begin{figure*}[t] 
    \centering
    \includegraphics[width=1\linewidth]{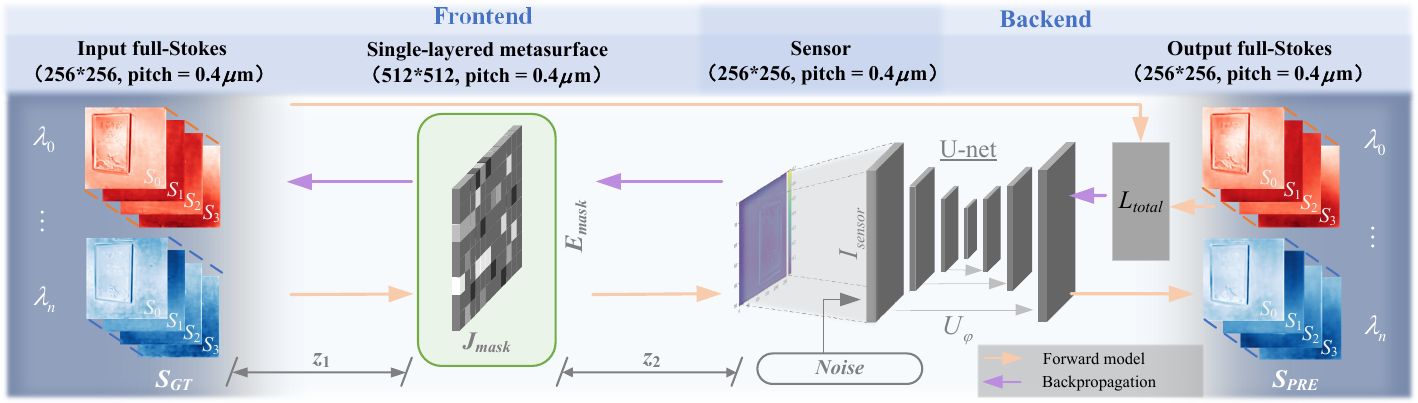}
    \caption{The architecture of the proposed full-Stokes polarization imaging system}
    \label{fig2}
\end{figure*}

The MLP-based surrogate model characterizes the responses of the subwavelength unit cells of the metasurface ($512\times 512$, pitch = 0.4 $\mathrm{\mu}$m) in the RGB-achromatic case\cite{li2022inverse, Pestourie2023}. As shown in Figure \ref{fig1}a, each unit cell consists of a $\text{TiO}_2$ rectangular nanopillar on a $\text{SiO}_2$ substrate \cite{sacchi2025silica}. The nanopillar height $h$ is fixed at 0.8 $\mathrm{\mu}$m for ease of fabrication, whereas the in-plane geometrical parameters, namely the length $a$, width $b$, and rotation angle $\theta$, are treated as degrees of freedom. The tunable parameters $a$ and $b$ are varied within the range of 0.05--0.35 $\mathrm{\mu}$m, and $\theta$ is swept from $0^{\circ}$ to $90^{\circ}$. The substrate height $d$ is fixed at 500 $\mathrm{\mu}$m, and the unit cell period $p$ is fixed at a subwavelength value of 0.4 $\mathrm{\mu}$m to suppress higher-order diffraction.

For each meta-atom, the local polarization response is described by a $2\times 2$ complex Jones matrix \cite{rubinjones}:
\begin{equation}
\bm{J}(a,b,\theta,\lambda)=
\bm{R}(-\theta)
\begin{bmatrix}
t_x(a,b,\lambda)e^{i\phi_x(a,b,\lambda)} & 0\\
0 & t_y(a,b,\lambda)e^{i\phi_y(a,b,\lambda)}
\end{bmatrix}
\bm{R}(\theta),
\end{equation}
where $t_x$ and $t_y$ denote the amplitude transmission coefficients along the nanopillar's principal axes, and $\phi_x$ and $\phi_y$ represent the corresponding phase shifts. These amplitude and phase responses are determined by the in-plane geometrical parameters $a$ and $b$, as well as the wavelength $\lambda$. The rotation matrices $\bm{R}(\theta)$ and $\bm{R}(-\theta)$ transform the principal-axis response of the anisotropic nanopillar into the global polarization basis, allowing the resulting Jones matrix elements to describe polarization-dependent amplitude modulation, phase modulation, and polarization conversion.

To generate the dataset for the surrogate model, we perform full-wave electromagnetic simulations using the finite-difference time-domain (FDTD) solver in MEEP under the locally periodic approximation (LPA)\cite{oskooi2010meep, Pestourie2018chromatic}. LPA is widely used in the design of conventional metasurfaces comprised of nanopillars or nanofins, in which non-local coupling between adjacent meta-atoms is negligible\cite{lin2019topology, Pestourie2018chromatic, Chung2023Inverse}. Two orthogonal linearly polarized incident fields, $[1,0]^{T}$ and $[0,1]^{T}$, are used as unit excitations to determine the complex Jones matrices in the RGB-achromatic case. The resulting dataset establishes the mapping between meta-atom geometry and polarization response.

The MLP-based surrogate model is then trained to approximate this mapping, as illustrated in Figure \ref{fig1}b. The input to the MLP is the set of geometrical parameters $\{a,b,\theta\}$, and the output is the real and imaginary parts of the Jones-matrix elements at the RGB wavelengths. Once trained, the surrogate model replaces the computationally expensive full-wave simulations during system-level optimization and enables efficient backpropagation through the optical frontend. The training loss is defined as the mean squared error (MSE) between the predicted and simulated Jones-matrix elements. The network is trained via custom gradient descent with a learning rate of $10^{-3}$ and an early stopping mechanism, using a training/validation/test data split of 8:1:1.

The prediction accuracy of the surrogate model is summarized in \textbf{Figure S1}. Strong agreement between simulated and predicted Jones-matrix elements is observed, with $R^{2}$ values (coefficient of determination) around 0.98 across the real and imaginary parts of $J_{xx}$, $J_{xy}$, $J_{yx}$, and $J_{yy}$ at the RGB wavelengths.

The entire forward pipeline can be formulated as follows, and it holds true for all the RGB wavelengths:

\begin{equation}
\begin{aligned}
\bm{E}_{\text{mask}}(\lambda_i) &= \bm{J}_{\text{mask}}(\lambda_i) \mathcal{A}_{z_{1}, \lambda_i}\left(\bm{E}_{\text{in}}(\lambda_i)\right),
\quad i = 1,2,\ldots,L, \\
I_{\text{sensor}} &= \sum_{i=1}^L \left|\mathcal{A}_{z_{2}, \lambda_i}\left(\bm{E}_{\text{mask}}(\lambda_i)\right)\right|^{2} + \eta
\end{aligned}
\end{equation}
Here, the distances from the metasurface to both the objective plane ($256 \times 256$, pitch = 0.4 $\mathrm{\mu}$m) and sensor plane ($256 \times 256$, pitch = 0.4 $\mathrm{\mu}$m) are $z_{1}$ and $z_{2}$. To maintain differentiability for end-to-end optimization, we use the vectorial angular spectrum propagation method to model the free space propagation of light, where $\mathcal{A}_{z_{1}, \lambda_i}$ and $\mathcal{A}_{z_{2}, \lambda_i}$ represent the angular spectrum propagation operators corresponding to $z_{1}$, $z_{2}$, and wavelength $\lambda_i$. The scene is represented by a spatially varying input field $\bm{E}_{\text{in}}$, whose amplitude and phase are derived from the target full-Stokes images. The field propagates through the metasurface represented by the Jones matrix $\bm{J}_{\text{mask}}$. $\bm{E}_{\text{mask}}$ is the output Jones vector field of the metasurface. The sensor measures only the scalar intensity $I_{\text{sensor}}$, which compresses the polarization and spectral information into a single channel. 
$\eta$ denotes the additive generic noise term, which is commonly modeled as zero-mean Gaussian white noise following the distribution \(\eta \sim \mathcal{N}(0,\sigma^2)\)\cite{sitzmann2018end}.

In the RGB-achromatic case, a $256 \times 256$ sensor measurement is used to reconstruct a $256 \times 256 \times 4 \times 3$ full-Stokes data cube (pitch = 0.4 $\mathrm{\mu}$m). Due to the incoherent nature of the imaging process, the sensor captures the linear summation of intensities, resulting in a high compression ratio of 12 in snapshot imaging. This makes the reconstruction strongly underdetermined and motivates joint optimization of both the optical frontend and the digital backend for better recovery.

\subsection{U-Net Backend}
\label{sec2.2}
\begin{figure}[!htbp] 
    \centering
    \includegraphics[width=0.7\textwidth]{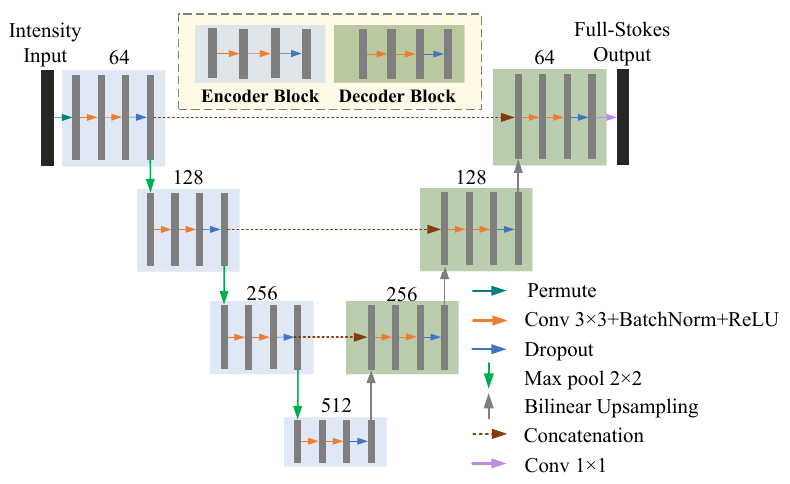}
    \caption{Schematic overview of the U-Net used in the backend. The values above the block represent the number of feature maps produced at each stage.}
    \label{unet}
\end{figure}

The polarization state of the incident field is represented by the full-Stokes parameters $S_{0}$, $S_{1}$, $S_{2}$, and $S_{3}$. Recovering these quantities from a single scalar intensity image constitutes a highly underdetermined inverse problem, particularly in the RGB-achromatic case. From the perspective of compressed sensing\cite{Duarte4472247}, this task can be regarded as a learned reconstruction problem from optically multiplexed low-dimensional measurements. To address this challenge, we employ a U-Net as the computational backend, as shown in \textbf{Figure \ref{unet}}.
The architecture of U-Net utilizes a symmetric, hierarchical encoder-decoder structure with skip-connections to fuse multi-scale spatial features, which is crucial for maintaining the fidelity of complex polarization edges and textures that are often blurred in standard auto-encoders. The encoder path progressively processes the input through a series of double convolution blocks with $3\times 3$ convolutional layers, batch normalization, ReLU activation, and dropout with a rate of 0.1, reaching a bottleneck with 512 channels to extract compact abstract representations of the encoded Stokes information. Symmetrically, the decoder path reconstructs the spatial resolution via bilinear upsampling while incorporating the encoder's high-frequency details through skip-connections. A $1\times 1$ convolutional layer is employed at the output to project the decoded features to the target channel dimension, which can be dynamically adjusted based on the number of wavelengths. Additionally, a global residual learning strategy is adopted where the network predicts the deviation from a baseline constructed from the input intensity, facilitating faster convergence during optimization. For each wavelength, the network directly predicts the Stokes parameters.

Notably, U-Net exhibits strong translation invariance via its convolutional architecture, aligning well with the approximately linear shift-invariant (LSI) propagation model of the imaging system\cite{ronneberger2015u}. This makes it possible to naturally learn the spatial mapping from sensor intensity measurements to full-Stokes polarization parameters, efficiently overcoming the limitations of handcrafted priors in traditional regularization methods. Moreover, U-Net supports end-to-end optimization with relatively higher computational efficiency than iterative optimization-based methods.

\subsection{End-to-end Collaboration}
\label{sec2.3}

The frontend and the backend are then jointly optimized on a multi-objective loss function $L_{\text{total}}$ that integrates the image reconstruction fidelity with the smoothness constraints to enhance manufacturability of the device. The collaborative refinements ensure high PSNR and SSIM performance even at a significant compression ratio. The total loss $L_{\text{total}}$ is defined as:
\begin{equation}
L_{\text{total}} = w_{\text{rec}}L_{\text{rec}} + w_{\text{ssim}}L_{\text{ssim}} + w_{\text{smooth}}L_{\text{smooth}}
\end{equation}
where $L_{\text{rec}}$ combines pixel-level $L_{1}$ and $L_{2}$ errors with a Sobel gradient loss to maintain high numerical accuracy and edge sharpness, $L_{\text{ssim}}$ constrains the structural integrity of the reconstructed Stokes fields and $L_{\text{smooth}}$ applies total variation (TV) regularization to the geometrical parameters to ensure spatial smoothness for feasible fabrication. The loss weights $w_{\text{rec}}$, $w_{\text{ssim}}$, $w_{\text{smooth}}$ are set to 10, 3, 0.01.

U-Net is trained jointly with the optical frontend using the Adam optimizer with a batch size of 4, and an early stopping mechanism. The dataset is split into training/validation/test sets with a ratio of 8.5:1:0.5. The checkpoint with the lowest validation loss is selected for testing.

Gradients from $L_{\text{total}}$ backpropagate through the entire differentiable pipeline, following a chain rule framework for all variables, including the metasurface geometry, the object/image distances, and the neural-network weights. The core gradient flow with respect to the metasurface geometric parameters is $\{a, b, \theta\}$ expressed as: 
\begin{equation}
\frac{\partial L_{\text {total }}}{\partial\{a, b, \theta\}}=\frac{\partial L_{\text {total }}}{\partial U_{\varphi}} \cdot \frac{\partial U_{\varphi}}{\partial I_{\text {sensor }}} \cdot \frac{\partial I_{\text {sensor }}}{\partial \bm{E}_{\text {mask }}} \cdot \frac{\partial \bm{E}_{\text {mask }}}{\partial \bm{J}_{\text {mask}}} \cdot \frac{\partial \bm{J}_{\text {mask}}}{\partial\{a, b, \theta\}}
\end{equation}
where $U_\varphi$ represents the U-Net decoder function.

This formulation allows the optical encoder and the neural decoder to be refined jointly toward the same objective. Rather than optimizing the optics for a predefined polarization response and training the network afterward, the proposed framework learns an optical encoding strategy that directly matches the reconstruction task and the statistics of the dataset.

\section{Results and Discussion}

We evaluate the proposed framework on a real-world spectro-polarimetric dataset \cite{jeon2024spectral}, which contains 2022 samples with substantially richer spatial, spectral, and polarization information as well as noise distributions from real-world sensors compared with the synthetic datasets commonly used in prior works \cite{xia2024joint}. To rigorously evaluate the reconstruction capability of the proposed framework, we have studied both monochromatic (Section \ref{sec3.1}) and RGB-achromatic reconstructions (Section \ref{sec3.2}) and compare four metasurface mask settings: a trainable mask, a fixed mask with an ideal lens profile, a random mask with random geometric parameters, and a no-mask baseline without the metasurface, where only the U-Net in the backend is optimized for the latter three cases. 1\,\% zero-mean Gaussian white noise is added to the normalized sensor intensity to emulate realistic measurement uncertainties and improve the robustness of the reconstruction network. The framework is trained on an NVIDIA GeForce RTX 5090. Meanwhile, the comparison between the proposed framework and the mainstream results is also conducted in Section \ref{sec3.3}.

\subsection{Monochromatic Reconstruction Evaluation}
\label{sec3.1}

\begin{figure}[htbp]
  \centering
  \begin{subfigure}{0.45\textwidth}
    \centering
    \includegraphics[width=\linewidth]{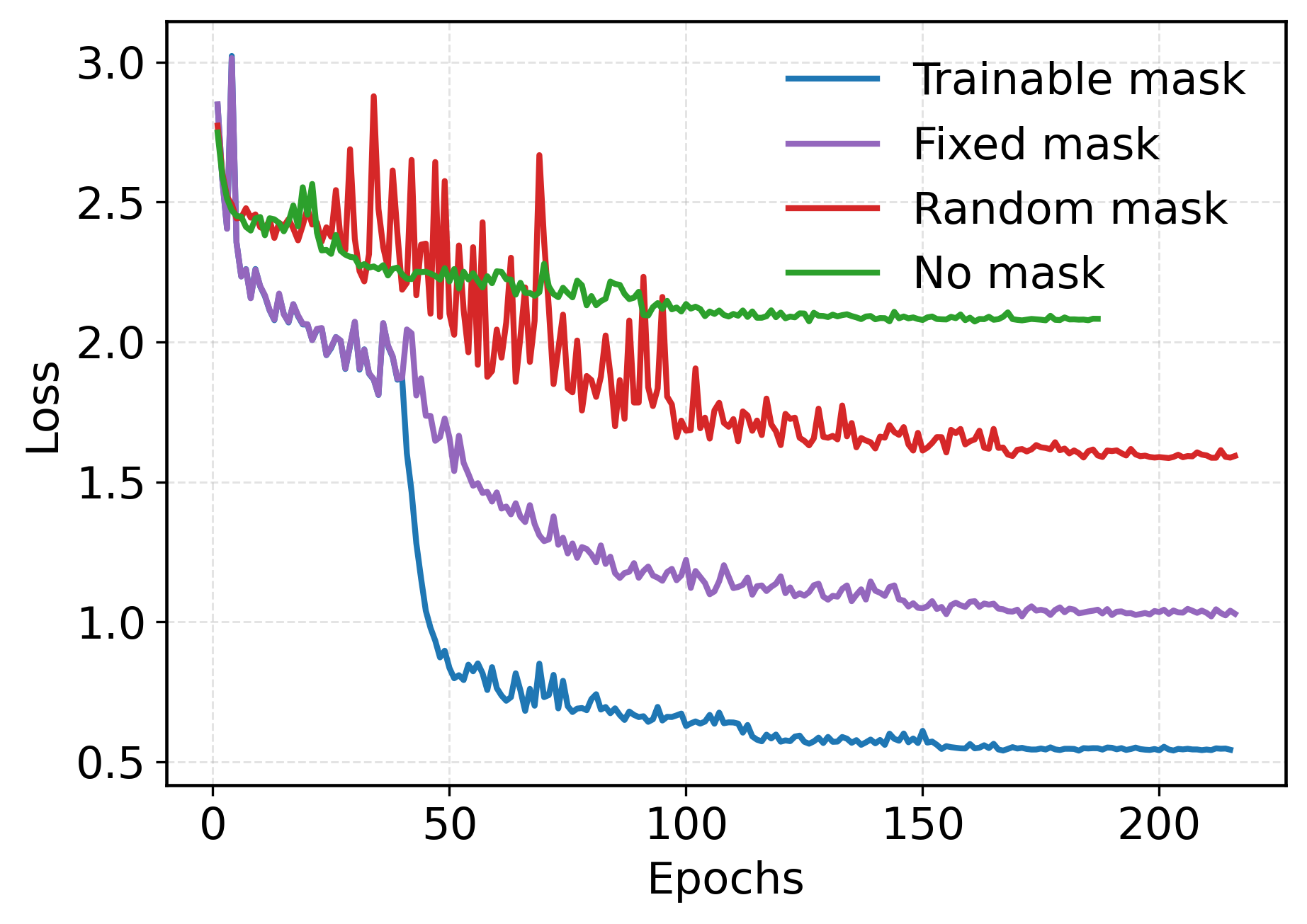}
    \caption{} 
    \label{fig9}
  \end{subfigure}
  \hfill 
  \begin{subfigure}{0.45\textwidth}
    \centering
    \includegraphics[width=\linewidth]{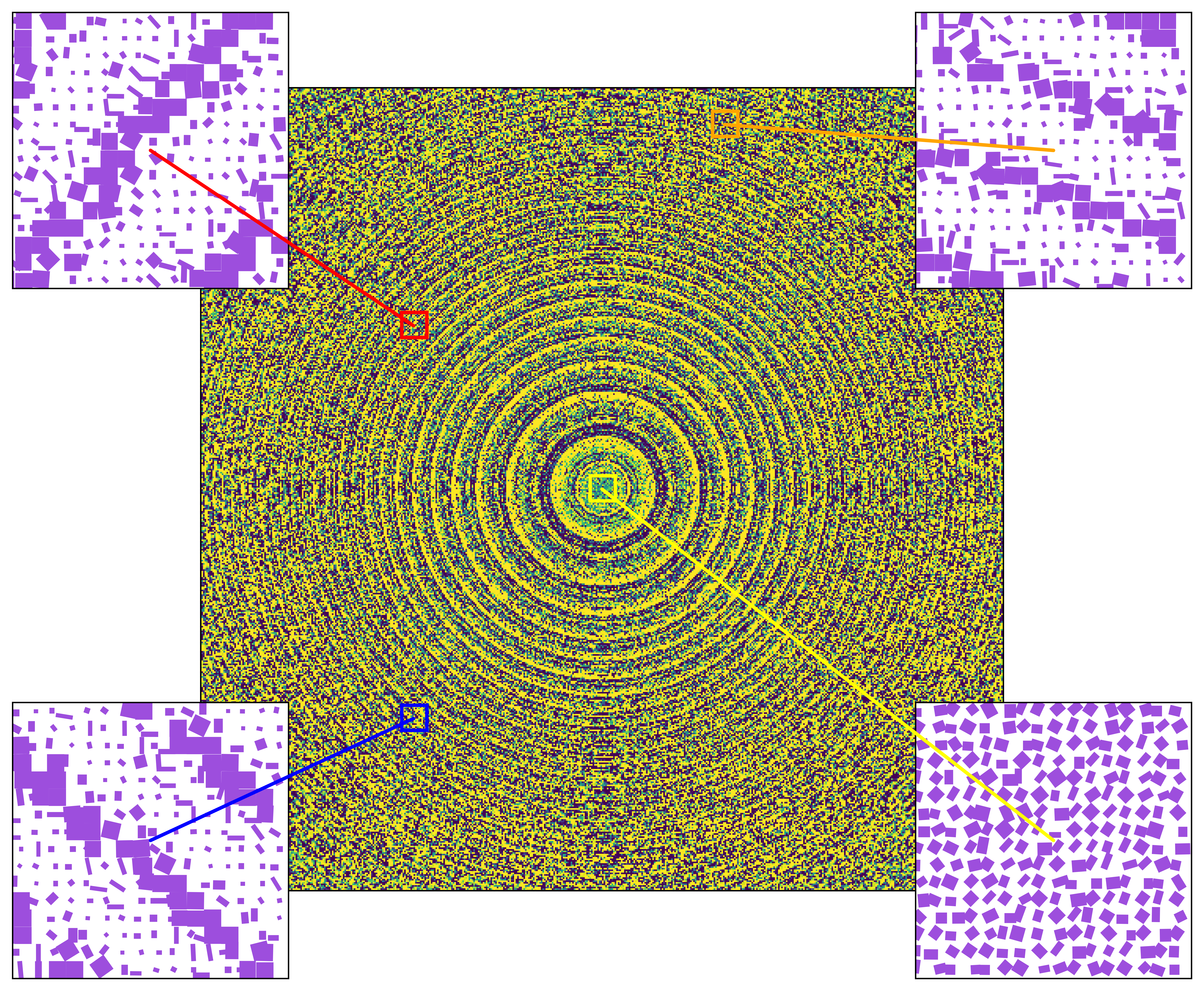}
    \caption{}
    \label{metaplt_single}
  \end{subfigure}
  \caption{(a) Loss function under different masks over training epochs in the monochromatic case. (b) The phase profiles of the trainable mask metasurface after end-to-end training in the monochromatic case.}
  \label{figures_Single}
\end{figure}

The monochromatic task encodes the full-Stokes parameters at a specific wavelength (0.44 $\mathrm{\mu}$m) into a single intensity image with a high compression ratio of 4. The monochromatic training dynamics are shown in \textbf{Figure \ref{figures_Single}a}. The trainable-mask configuration converges to a lower validation loss ($\approx 0.54$) in 215 epochs within 2.40 h, compared with the fixed-mask ($\approx 1.03$), random-mask ($\approx 1.59$) and no-mask  ($\approx 2.08$) baselines which are early-stopped at 216, 216 and 187 epochs. This demonstrates that joint optical-digital optimization improves the conditioning of the inverse problem. In other words, the optimized metasurface produces sensor measurements that capture more information for downstream Stokes recovery than either an ideal lens encoding, random encoding or no encoding at all. The validation loss of the trainable mask system is consistent with the residual phase mismatch, aperture discretization, and optical nonidealities, making the forward operator hard to invert. 

The optimized phase profile of the trainable mask is shown in Figure \ref{figures_Single}b. Without the explicit metalens constraint, the learned phase distribution spontaneously evolves into a lens-like profile with fine tuning, supporting the spatial imaging with polarization modulations. Such a phase profile naturally supports image formation and yields an approximately LSI imaging channel\cite{khorasaninejad2016metalenses, chen2018a, limeta, khorasaninejad2016polarization, Sun2025}. This property is particularly advantageous for the U-Net reconstruction backend, whose convolutional architecture implicitly assumes spatially shared image features and local translation equivariance\cite{ronneberger2015u}.

\begin{figure}[t] 
    \centering
    \includegraphics[width=0.7\textwidth]{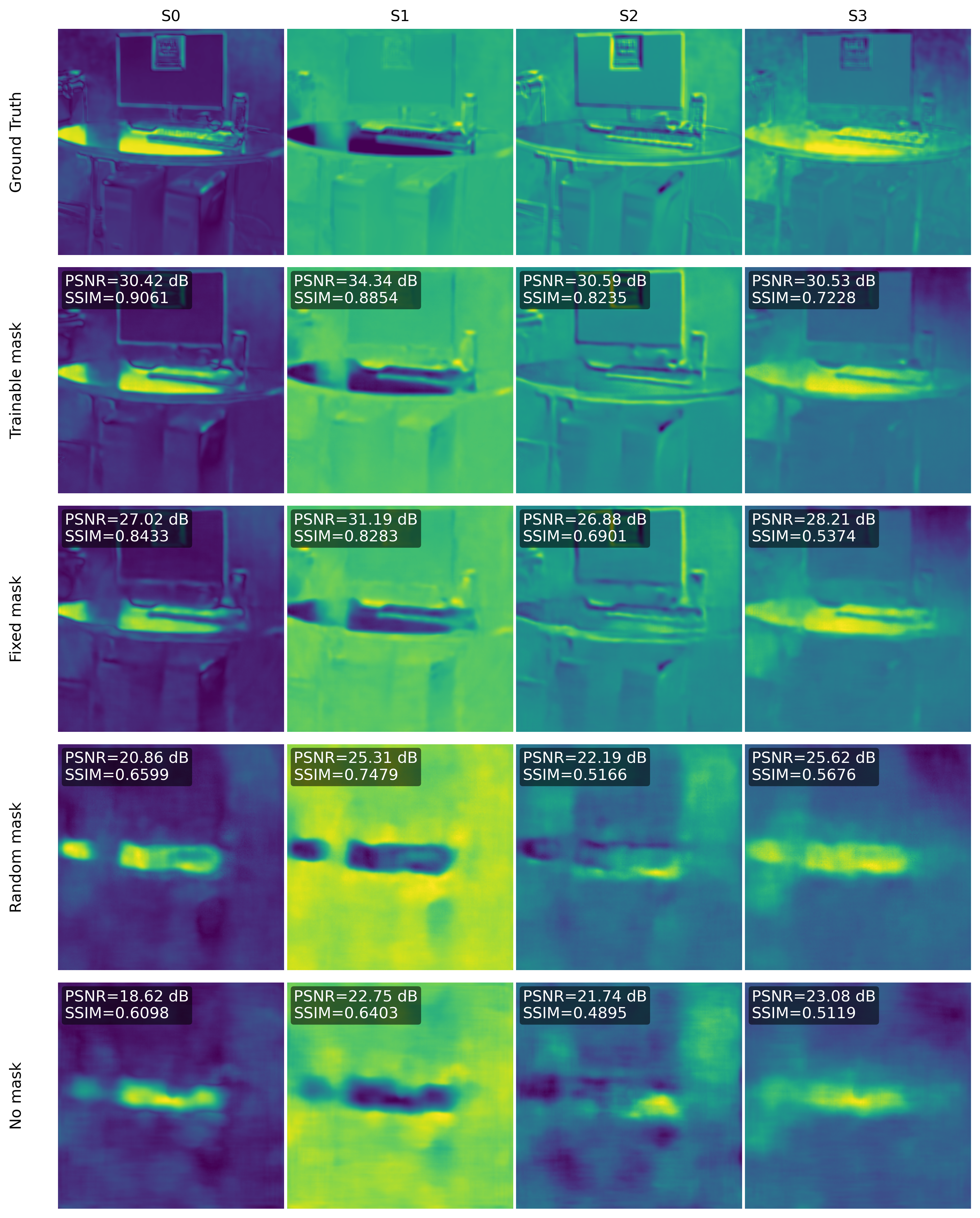}
    \caption{Examples of full-Stokes images reconstructed under different masks in the monochromatic case with 1\,\% zero-mean Gaussian white noise.}
    \label{fig10}
\end{figure}

\begin{table}[t]
\caption{Quantitative comparison of PSNR, SSIM, MSE and inference time under different masks in the monochromatic case with 1\,\% zero-mean Gaussian white noise.}
\centering
\begin{tabular}{lllll}
\toprule
Methods & Trainable mask & Fixed mask & Random mask & No mask \\
\midrule
PSNR (dB) & 27.06 & 23.56 & 19.63 & 18.25 \\
SSIM & 0.7172 & 0.5664 & 0.3959 & 0.2964 \\
MSE & $6.86 \times 10^{-4}$ &$1.69 \times 10^{-3}$ & $4.71 \times 10^{-3}$ & $6.29\times 10^{-3}$ \\
Inference time (ms/img) & 2.46 & 2.45 & 2.47 & 2.49 \\
\bottomrule
\end{tabular}
\label{tab:mono_stage2}
\end{table}
The quantitative performance is listed in \textbf{Table \ref{tab:mono_stage2}}. The trainable mask configuration achieves an average PSNR of 27.06 dB and an average SSIM of 0.7172, compared with 23.56 dB and 0.5664 for the fixed-mask baseline, 19.63 dB and 0.3959 for the random-mask baseline and 18.25 dB and 0.2964 for the no-mask baseline. This corresponds to a PSNR gain of 3.50 dB over the fixed-mask case, 7.43 dB over the random-mask case and 8.81 dB over the no-mask case, together with substantial SSIM improvements of > 0.15. The MSE is reduced from $1.69 \times 10^{-3}$, $4.71 \times 10^{-3}$ and $6.29 \times 10^{-3}$ in the three baselines to $6.86 \times 10^{-4}$ in the trainable mask setting. Given the fast transmission speed of the physical frontend, we only evaluate the inference time in the backend, which remains essentially unchanged at approximately 2.5 ms/img and thus indicates ultrafast snapshot imaging.

Representative reconstructions are shown in \textbf{Figure \ref{fig10}}. The trainable mask configuration more accurately preserves fine structures and spatial boundaries in the reconstructed Stokes images, whereas the fixed-mask, random-mask and no-mask baselines exhibit blurrier edges and a higher loss of local polarization detail. These results confirm that end-to-end optimization yields a measurable advantage over fixed, random encoding and purely computational reconstruction.

\begin{figure}[htbp]
  \centering
  \begin{subfigure}{0.45\textwidth}
    \centering
    \includegraphics[width=\linewidth]{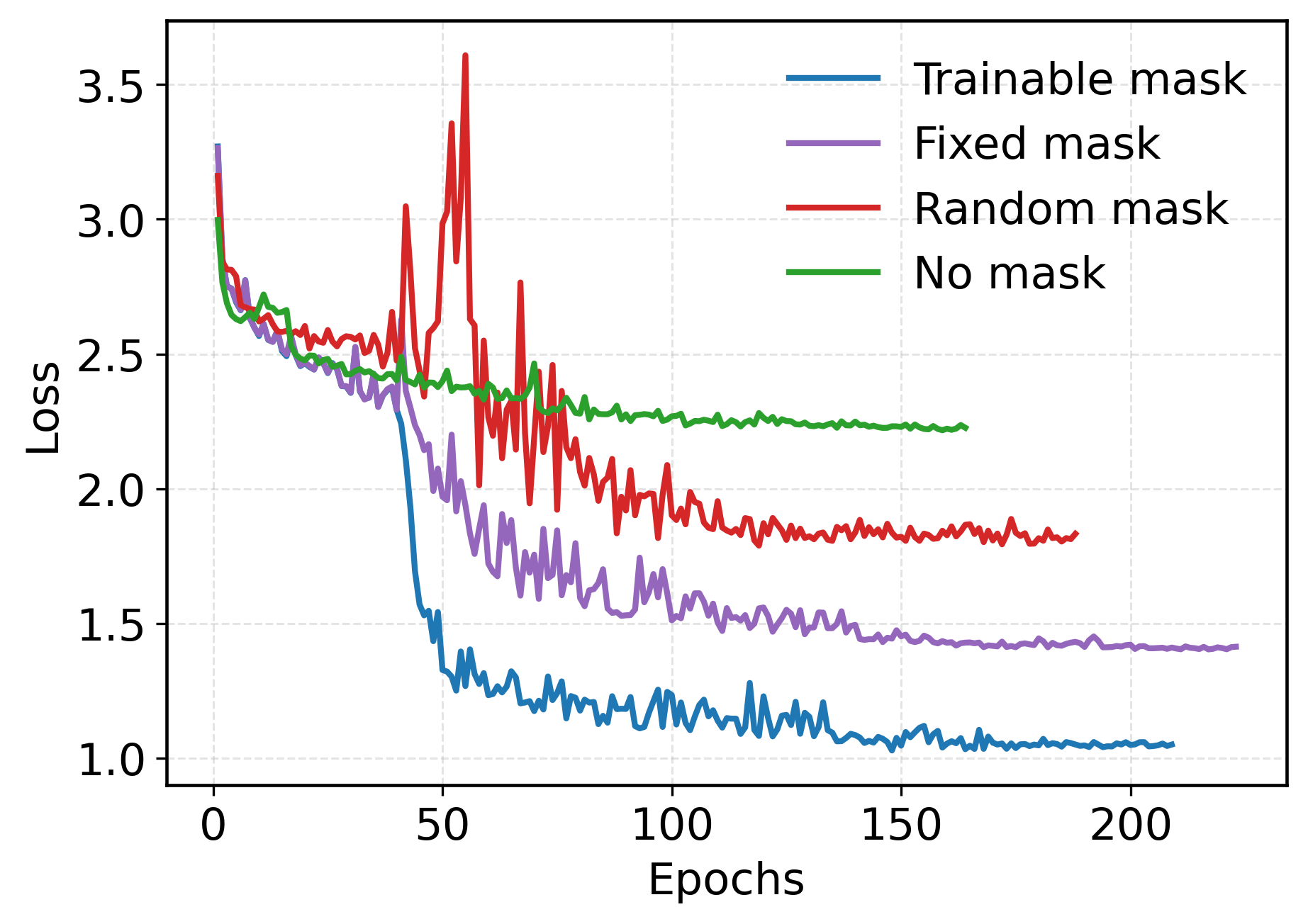}
    \caption{} 
    \label{fig11}
  \end{subfigure}
  \hfill 
  \begin{subfigure}{0.45\textwidth}
    \centering
    \includegraphics[width=\linewidth]{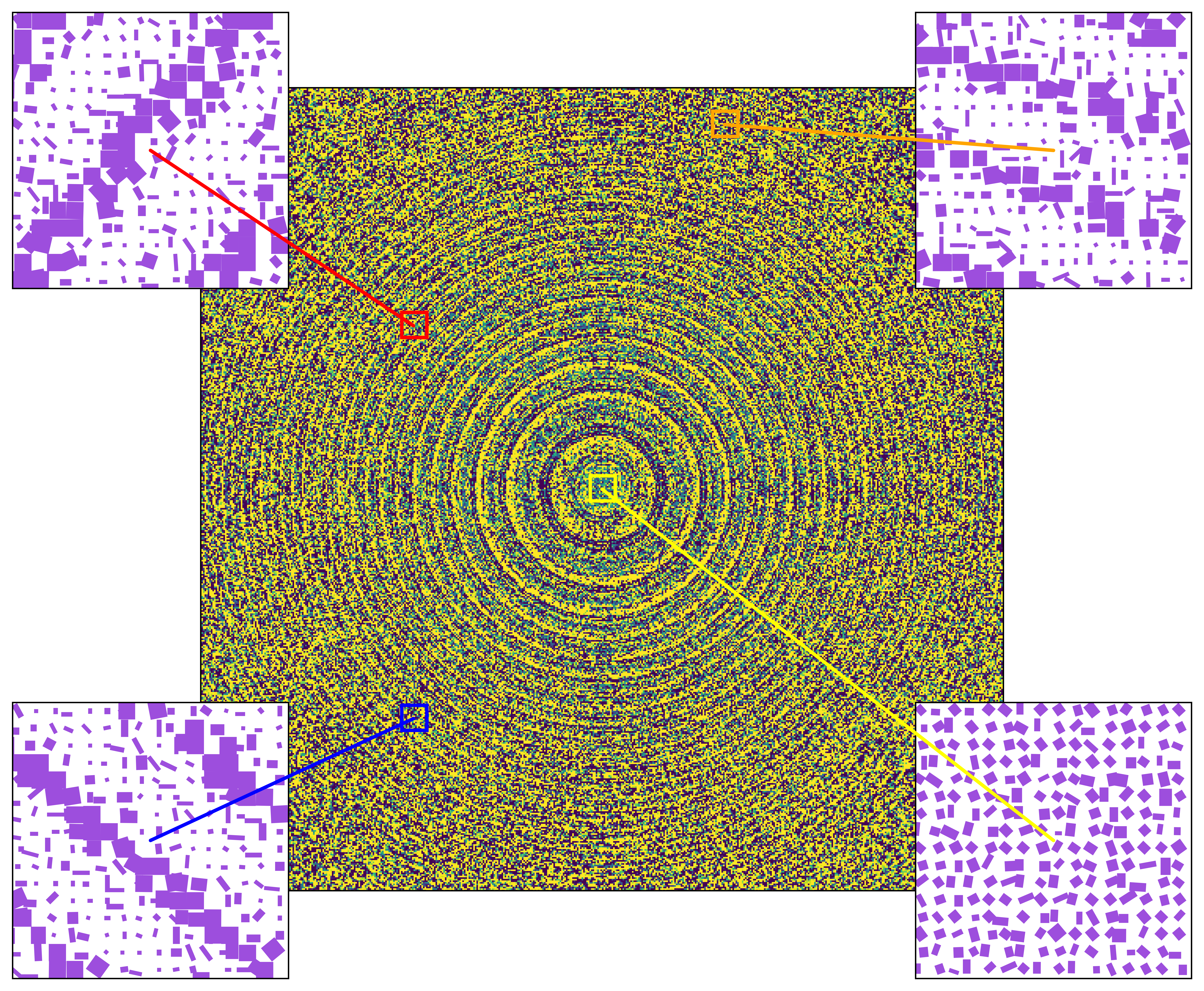}
    \caption{}
    \label{metaplt_RGB}
  \end{subfigure}
  \caption{(a) Loss function under different masks over training epochs in the RGB-achromatic case. (b) The phase profiles of the trainable mask metasurface after end-to-end training in the RGB-achromatic case.}
  \label{figures_RGB}
\end{figure}

\begin{figure}[t] 
    \centering
    \includegraphics[width=0.7\textwidth]{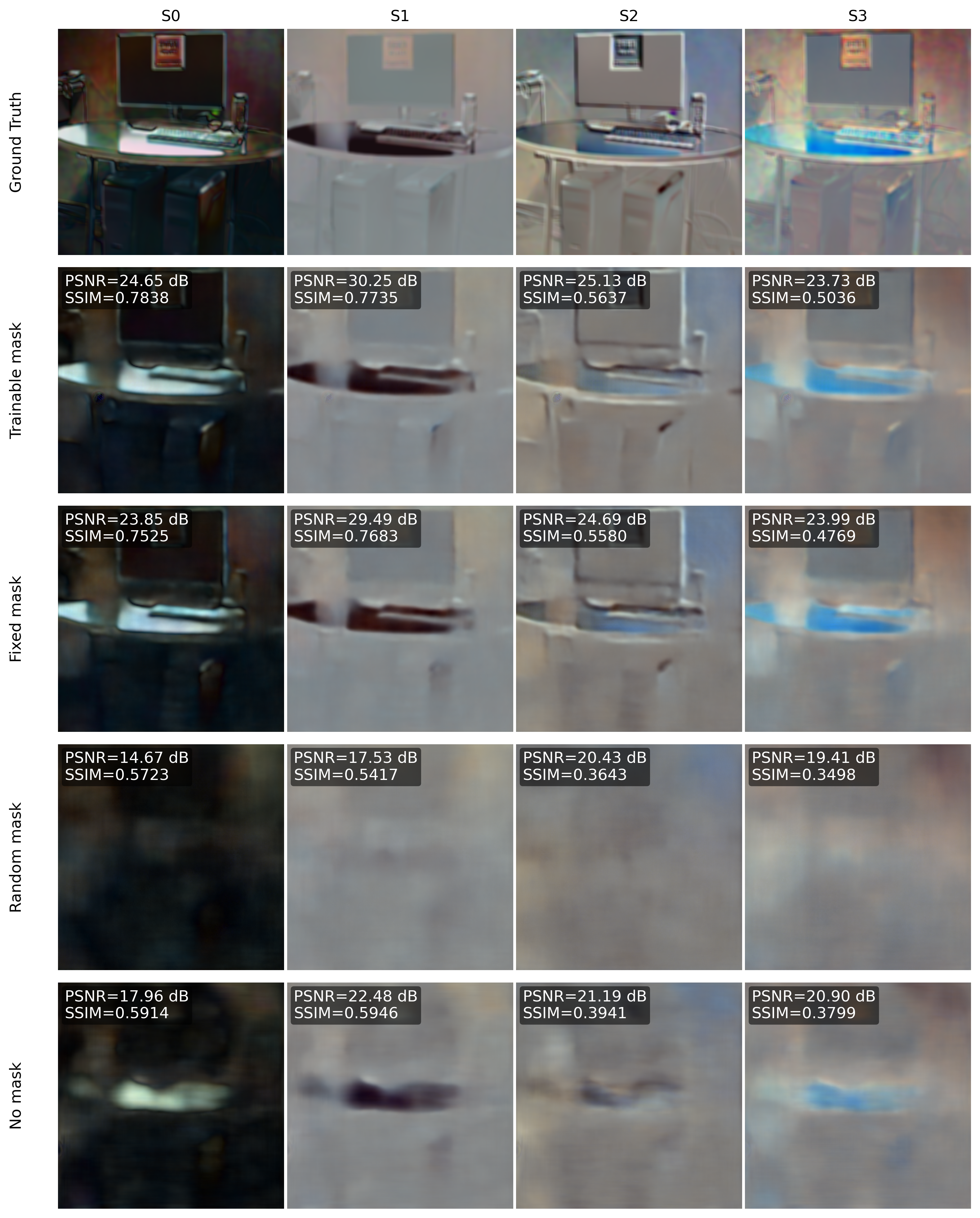}
    \caption{Examples of full-Stokes images reconstructed under different masks in the RGB-achromatic case with 1\,\% zero-mean Gaussian white noise.}
    \label{fig12}
\end{figure}

\subsection{RGB-achromatic Reconstruction Evaluation}
\label{sec3.2}

The RGB reconstruction task is significantly more challenging because the sensor must encode the full-Stokes parameters at three different wavelengths into a single intensity image with an even higher compression ratio of 12. The training results are shown in \textbf{Figure \ref{figures_RGB}a}. The validation loss for the trainable-mask configuration drops sharply around epoch 40 and keeps decreasing steadily to approximately 1.05 at around epoch 209 with a training time of 5.72 h, compared with the fixed-mask ($\approx 1.41$), random-mask ($\approx 1.83$) and no-mask ($\approx 2.23$) baselines which are early-stopped at 223, 188 and 164 epochs. The trainable-mask configuration achieves a relatively higher validation loss than in the monochromatic case, because the RGB task is more severely underdetermined. The optimized phase profile of the trainable-mask metasurface is shown in Figure \ref{figures_RGB}b. It is shown that the optimization can automatically discover the lens-like profile even in the RGB-achromatic case.

\begin{table}[t]
\caption{Quantitative comparison of PSNR, SSIM, MSE and inference time under different masks in the RGB-achromatic case with 1\,\% zero-mean Gaussian white noise.}
\centering
\begin{tabular}{lllll}
\toprule
Methods & Trainable mask & Fixed mask & Random mask & No mask \\
\midrule
PSNR (dB) & 23.35 & 21.33 & 18.72 & 18.03 \\
SSIM & 0.5643 & 0.4743 & 0.3620 & 0.2775 \\
MSE & $1.87 \times 10^{-3}$ & $3.07 \times 10^{-3}$ & $5.88 \times 10^{-3}$ & $6.90 \times 10^{-3}$ \\
Inference time (ms/img) & 2.49 & 2.47 & 2.51 & 2.45 \\
\bottomrule
\end{tabular}
\label{tab:rgb_stage2}
\end{table}

The meta-optic system with trainable mask reaches an average PSNR of 23.35 dB and an average SSIM of 0.5643 at the RGB wavelengths (\textbf{Table \ref{tab:rgb_stage2}}), compared with 21.33 dB and 0.4743 for the fixed-mask baseline, 18.72 dB and 0.3620 for the random-mask baseline and 18.03 dB and 0.2775 for the no-mask baseline. This corresponds to PSNR gains of 2.02 dB, 4.63 dB and 5.32 dB, respectively, together with marked SSIM improvements of >0.09. The MSE of the trainable mask is around $1.87 \times 10^{-3}$, which remains lower than those of the baselines. The inference time remains close to 2.5 ms/img, enabling real-time snapshot imaging and demonstrating that the RGB-achromatic configuration does not sacrifice the speed of the neural decoder. 

The visual results in \textbf{Figure \ref{fig12}} show that the proposed system maintains good reconstruction fidelity across different wavelength channels and Stokes components despite using a compact single-layered meta-optic system with a high compression ratio of 12. Compared with the baselines, the trainable mask better preserves both spectral consistency and polarization-dependent contrast. This result is important because it shows that the benefit of end-to-end optical design is not limited to the monochromatic case but extends to high-dimensional RGB full-Stokes recovery.

\subsection{Full-Stokes Reconstruction Performance Comparison}
\label{sec3.3}
To place the results in context, \textbf{Table \ref{tab:sota_comparison}} compares the proposed framework with representative full-Stokes reconstruction systems reported in the literature. Direct numerical comparison should be interpreted with caution because the reported methods differ in wavelength range, compression ratio, dataset type and meta-optic architecture. The type of dataset can be divided into real-world, experimental, and synthetic. The real-world dataset is physically captured from practical scenes where the target information arises from native interactions between material, illumination and geometry rather than prescribed labels or staged scene design. The experimental dataset is physically captured in a staged laboratory setup with deliberately designed targets, illumination, polarization states, scattering media, depths, or acquisition conditions. The synthetic dataset is generated by simulation, rendering, or computational construction rather than direct physical acquisition of the quantity of interest. In particular, many prior works are evaluated on experimental datasets \cite{ren2025single} or synthetic datasets \cite{xia2024joint, xia2025inverse}, whereas our results are obtained on a real-world spectro-polarimetric dataset containing more complex scenes with real-world polarization information. Meanwhile, the optical systems in many prior works comprised multiple bulky optical elements including diffuser, lens, bandpass filter, polarizer and quarter-waveplate aside from the metasurfaces \cite{ren2025single, xia2024joint, xia2025inverse}, whereas our system consists of a single-layered metasurface alone with a much more compact footprint.

On a real-world spectro-polarimetric dataset\cite{jeon2024spectral}, the proposed framework remains competitive while operating under aggressive compression with a single-layered metasurface. The monochromatic configuration achieves a high compression ratio of 4 with 27.06 dB PSNR and 0.7172 SSIM, while the RGB configuration achieves an even more aggressive compression ratio of 12 with 23.35 dB PSNR and 0.5643 SSIM. These results show that the proposed end-to-end framework can maintain high reconstruction fidelity in an ultra-compact footprint even when both spectral and polarization information are significantly compressed. The performance is sufficient for many applications such as classification and pattern recognition\cite{Goudail:04, Hsing:26}. 

To test the generalizability of the proposed framework, it has been evaluated on the same experimental dataset introduced in \cite{ren2025single}. Our framework achieves 34.17 dB PSNR and 0.9076 SSIM. Although this result is lower than the 41.71 dB PSNR reported in \cite{ren2025single}, it still indicates high-fidelity reconstruction, especially considering that our design has a high compression ratio of 4 with $\sim 62$-fold improvement.

\begin{table*}[h]
\caption{Comparison between the proposed framework and the mainstream snapshot full-Stokes reconstruction works.}
\centering
\small 
\setlength{\tabcolsep}{4pt} 
\begin{tabular}{cccccccc} 
\toprule
Wavelength (nm) & Compression ratio & Dataset & Meta-optic architecture & Ref. \\
\midrule
808 & 0.0642 & experimental & single-layered metasurface+diffuser & \cite{ren2025single} \\
435.8 & 4 & experimental\cite{ren2025single} & single-layered metasurface & Ours \\
530 & 4 & synthetic & \makecell{cascaded metasurfaces+five conventional optical elements} & \cite{xia2025inverse} \\
532 & 4 & synthetic & double-layered metasurface+lens & \cite{xia2024joint} \\
435.8 & 4 & \textbf{real-world} & \textbf{single-layered metasurface} & Ours \\
435.8, 546.1, 700 & \textbf{12} & \textbf{real-world} & \textbf{single-layered metasurface} & Ours \\
\bottomrule
\end{tabular}

\label{tab:sota_comparison}
\end{table*}

\section{Summary and Outlook}

In summary, we have presented an end-to-end framework for snapshot full-Stokes polarization imaging. The single-layered meta-optic frontend and a neural reconstruction backend are jointly optimized. By employing a trainable MLP-based surrogate model, the differentiable meta-optics frontend learns an optical measurement operator tailored for full-Stokes polarization recovery rather than intensity-only imaging. Notably, on a real-world dataset, the compact single-layered metasurface reconstructs monochromatic and RGB-achromatic full-Stokes images from a snapshot measurement, with the RGB configuration operating at an aggressively high compression ratio of 12.

Beyond the absolute reconstruction metrics, the central result is that co-designed optical encoding substantially improves the reconstructability of the inverse problem relative to fixed or random encoders. The main reason for the performance gain does not arise merely from a stronger reconstruction network, but from the joint learning process of the task-specific optical encoder and the digital decoder. This is critical for RGB-achromatic full-Stokes polarization imaging, where spectral and polarization information are compressed into a single information channel under realistic physical constraints. This result supports the central premise of end-to-end meta-optical imaging: for highly compressed inverse problems, the quality of the frontend optical encoding is as important as the expressive power of the reconstruction backend. Additionally, the meta-optic system designed by the proposed framework is experimentally feasible by incorporating rectangular nanopillars.

Looking forward, future work should focus on more fabrication-robust and sensor-aware system designs that are compatible with practical pixel sizes and more expressive optical devices such as multilayer or nonlocal metasurfaces\cite{Stefanini2024, zhou2018multilayer, yang2025nonlocal}. Extending the present framework from three isolated wavelengths to continuous spectral bands\cite{colburn2018metasurface, Huang2024broadband}, from narrow-field operation to wide-FOV imaging\cite{hongli2024broadband, javed2026nearly, yeo2025neural, wirth2025wide}, from diffraction-limited systems to super-resolution imaging\cite{zhang2026supreso, tian2011survey} and from two-dimensional polarization recovery to depth-resolved polarimetric imaging \cite{arya2024end} could further broaden its applicability in scenarios such as compact machine vision\cite{liang2024metasurface}, remote sensing\cite{tyo2006review}, biomedical imaging\cite{wang2024use}, and physics-informed imaging\cite{lin2022end}. Within that broader trajectory, the present work shows that meta-optics can be used not only as ultrathin replacements for conventional optical elements, but as trainable physical encoders whose function is defined directly by these downstream inference tasks. We therefore expect end-to-end optical–digital co-design to become an increasingly important route toward application-ready imaging systems.

\medskip
\textbf{Supporting Information} \par 
Supporting Information is available from the Wiley Online Library or from the author.
\begin{itemize}
\item Metasurface MLP-based surrogate model performance. (metasurface\_MLP\_surrogate\_performance.pdf)
\item Optimized metasurface parameters in the monochromatic case. (metasurface\_parameters\_RGB.xlsx)
\item Optimized metasurface parameters in the RGB-achromatic case. (metasurface\_parameters\_single.xlsx)
\end{itemize}

\medskip
\textbf{Funding} \par 
Advanced Materials-National Science and Technology Major Project (2025ZD0616302); the National Natural Science Foundation of China (62275046); the Jiangsu Province Frontier Research Project (BF2024061); the Jiangsu Key Laboratory for Undersea Communications and Sensing (8506006260C); the Big Data Computing Center of Southeast University and the Center for Fundamental and Interdisciplinary Sciences of Southeast University. 

\medskip
\textbf{Acknowledgements} \par 
The authors would like to thank Zin Lin for informative discussions.

\medskip


\bibliographystyle{MSP}
\bibliography{sample}



\end{document}